\documentstyle[12pt]{article}
\title{Mesoscopic Casimir forces from effects
of discrete particle number in the quantum vacuum.}
\author{ G.E. Volovik\\
Low Temperature Laboratory,
Helsinki University of Technology\\
P.O.Box 2200, FIN-02015 HUT, Finland\\
and\\
L.D. Landau Institute for Theoretical Physics, \\
Kosygin Str. 2, 117940 Moscow, Russia\\
}
\begin{document}

\maketitle
\vskip 1 truecm

\begin{abstract}
{Traditionally it is assumed that the Casimir vacuum
pressure does not depend on the ultraviolet cut-off. There
are, however, some arguments that the effect
actually depends on the regularization procedure and thus
on the trans-Planckian physics. We provide the condensed
matter example where the Casimir forces do explicitly
depend on the microscopic (correspondingly trans-Planckian)
physics due to the mesoscopic finite-$N$ effects, where
$N$ is the number of bare particles in condensed matter (or
correspondingly the number of the elements comprising the
quantum vacuum). The finite-$N$ effects lead to mesoscopic
fluctuations of the vacuum pressure. The amplitude of the
mesoscopic flustuations of the Casimir force in a system
with linear dimension
$L$ is by the factor
$N^{1/3}\sim L/ a_{\rm Planck} $ larger than the
traditional value of the Casimir force given
by effective theory, where $a_{\rm Planck}=\hbar/p_{\rm
Planck}$ is the interatomic distance which plays the
role of the Planck length.}

\end{abstract}

\section{Introduction.}

The attractive force between two parallel metallic plates in vacuum
induced by the vacuum fluctuations of the electromagnetic field has been
predicted by Casimir in 1948 \cite{Casimir}.  The calculation of the
vacuum presure is based on the regularization schemes, which allows to
separate the effect of the low-energy modes of the vacuum
from the huge diverging contribution of the high-energy degrees of the
freedom. There are different regularization schemes:
Riemann's zeta-function regularization; introduction of the
exponential cutoff; dimensional regularization, etc.
People are happy when different regularization schemes give
the same results. But this is not always so (see e.g.
\cite{Kamenshchik,Ravndal,Falomir}, and in particular
the divergencies occurring for spherical geometry in odd
spatial dimension are not
cancelled \cite{Milton,CasimirForSphere}). This raises
some criticism against the regularization methods
\cite{Hagen} or even some doubts concerning the existence
and the magnitude of the Casimir effect.

The same type of the Casimir effect arises in condensed matter, due to
thermal (see review paper \cite{Kardar}) or/and quantum
fluctuations.  When consindering the analog of the Casimir
effect in condensed matter,  the following correspondence
must be taken into account.   The ground state of quantum
liquid corresponds to the vacuum of quantum field theory.
The low-energy bosonic and fermionic quasiparticles in
quantum liquid correspond to matter. The low energy modes
with linear spectrum
$\omega=c_sp$ can be described by the
relativistic-type effective theory. The speed of sound $c_s$
or of other collective modes (spin waves, etc.) plays the
role of the speed of light. This ``speed of light'' is the
``fundamental constant'' which enters the effective theory
(quantum hydrodynamics in quantum liquids, or
electromagnetic theory in real vacuum). The fundamental
constants of the effective theory can be in principle
calculated using the microscopic physics, an analog of the
transPlanckian physics. The effective theory is valid only
at low energy, which is much smaller than the ``Planck
cutoff''. In quantum liquids the analog of the Planck energy
scale
$E_{\rm Planck}$ is determined either by
the mass
$m$ of the atom of the liquid, $E_{\rm Planck}\equiv
mc_s^2$, or by the Debye energy, $E_{\rm Planck}\equiv
\hbar c_s/a_{\rm Planck}$, where  $a_{\rm Planck}$ is
the interatomic distance which plays the role of the Planck
length  \cite{PhysRepRev}.

In some cases the analogy between effective
theories in quantum vacuum and in quantum liquids becomes
exact. For example, the low-energy fermionic and bosonic
collective modes can correspond to the chiral fermions, and
gravitational and gauge fields.  This allows to simulate
in condensed matter such phenomena as chiral
anomaly, event horizon, etc., (see
review \cite{PhysRepRev}).

The advantage of the quantum liquid is that the structure of the quantum
vacuum is known at least in principle. That is why one
can  calculate everything starting from the first principle
microscopic theory. For example, one can calculate the
vacuum energy under different external conditions, without
invoking any cut-off or regularization scheme.  Then one
can compare the results with what can be obtained within
the effective theory which deals only with the low-energy
phenomena. The latter requires the regularization scheme in
order to cancel the ultraviolet divergency, and thus one
can judge whether and which of regularization schemes are
physically relevant.

The traditional Casimir effects deals with the low energy massless modes.
The typical massless modes in quantum liquid are
sound waves. The acoustic field is desribed by the effective theory and
corresponds to the massless scalar field. The walls provide the boundary
conditions for the sound wave mode, usually these are the Neumann boundary
conditions. Because of the quantum hydrodynamic fluctuations there must be
the Casimir force between two parallel plates immersed in the quantum
liquid. Within the effective theory the Casimir force is given by the same
equation as the Casimir force acting between the conducting walls due to
quantum electromagnetic fluctuations. The only modifications are: (i)
the speed of light must be substututed by the spin of sound $c_s$;  (ii)  the
factor $1/2$ must be added, since we have the scalar field of the
longitudinal sound wave instead of two polarizations of light. If
$a$ is the distance between the plates and $A$ is their area, then the
$a$-dependent contribution to the ground state energy of the quantum liquid
at $T=0$ which follows from the effective theory must be
\begin{equation}
E_C= -{\hbar c_s\pi^2 A\over 1440 a^3}
\label{CasimirForceSound}
\end{equation}
Such microscopic quantities of the quantum liquid as the
mass of the atom $m$ and interatomic space $a_{\rm Planck}$
do not enter explicitly the Eq.(\ref{CasimirForceSound}):
the traditional Casimir force is completely determined by
the ``fundamental'' parameter $c_s$ of the effective
scalar field theory.

However, we shall show that the Eq.(\ref{CasimirForceSound})
is not always true. We shall give here an example, where
the effective theory is not able to predict the Casimir
force,  since the microscopic high-energy degrees of
freedom become important. In other words the
``transPlanckian physics'' shows up and the ``Planck''
energy scale explicitly enters the result. In this situation
the Planck scale is physical and cannot be removed by
any regularization.

The Eq.(\ref{CasimirForceSound}) gives a finite-size
contribution to the energy of quantum liquid. It is
inversly proportional to the linear dimension of the sistem,
$E_C \propto 1/L$. However, for us it is important that it
is not only the finite-size effect, but also the
finite-$N$ effect, $ E_C
\propto N^{-1/3}$, where
$N$ is the number of atoms in the liquid in the slab,
which is a discrete quantity.
Since
the main contribution to the vacuum energy is
$\propto L^3 \propto N$, the
relative correction of order $N^{-4/3}$ means that the
Casimir force is the mesoscopic effect. We shall show that
in quantum liquids, the essentially larger mesoscopic
effects, of the relative order $N^{-1}$, can be more
pronounced.  Such a finite-$N$ effect cannot be described
by the effective theory dealing
with the continuous medium, even if the theory includes the
real boundary conditions with the frequency dependence of
dielectic permeability.

We shall start with the simplest quantum ``liquid'' -- the
one-dimensional Fermi gas -- where the mesoscopic Casimir
forces can be calculated exactly without invoking any
regularization procedure.

\section{Vacuum energy and cosmological constant.}

\subsection{Vacuum energy from microscopic theory}

We consider the system of $N$ one-dimensional massless fermions, whose
continuous energy spectrum is
$\omega(p)=cp$, with $c$ playing the role of speed of
light. Let us start with the microscopic theory, which is
extremely simple: at
$T=0$ the fermions simply occupy all the energy levels
below the chemical potential $\mu$.   In the continuous
limit the total number of particles $N$ and the total energy
in the one-dimensional ``cavity'' of size $a$ are expressed
in terms of the Fermi momentum $p_F=\mu/c$ in the following
way
\begin{eqnarray}
N=na=a\int_{-p_F}^ {p_F}{dp\over 2\pi\hbar} ={ap_F\over \pi\hbar}~,
\label{FermiMomentum1}\\
E=\epsilon(n) a=a\int_{-p_F}^ {p_F}{dp\over 2\pi\hbar}cp ={acp_F^2\over
2\pi\hbar}={\pi\over 2}  \hbar c a n^2~.
\label{FermiMomentum2}
\end{eqnarray}
Here $n$ is the particle density. The  vacuum energy density of this
condensed matter as a function of $n$ is
$\epsilon(n)=(\pi
\hbar c/2) n^2$. The equation of state  comes from the thermodynamic
identity relating the pressure $P$ and the energy:
\begin{equation}
P=\mu n-
\epsilon~,
\label{ThermodIdentity}
\end{equation}
 where $\mu= d\epsilon/dn$ $=cp_F$ is the chemical potential. In our
case  $\mu=cp_F$ and and obtains the equation of state for our vacuum
\begin{equation}
P= \epsilon
\label{EquationState}
\end{equation}
which is conventional for
the system of 1+1 relativistic fermions.

\subsection{Vacuum energy in effective theory}

As distinct from the microscopic theory, which deals with bare particles,
the effective theory deals with the  quasiparticles -- fermions living at the
level  of the chemical potential $\mu=cp_F$. There are 4 different
quasiparticles: (i) quasiparticles and quasiholes living in the vicinity of
the Fermi point
$p_z=+p_F$ have spectrum $\omega_{qp}(p_+)=|\omega(p) -\mu|=c|p_+|$,
where $p_+=p_z-p_F$;   (ii) quasiparticles and quasiholes living in the
vicinity of the other Fermi point at
$p_z=-p_F$ have the spectrum $\omega_{qp}(p_-)=|\omega(p)
-\mu|= c|p_-|$,  where $p_-=p_z+p_F$. In the effective
theory the energy of the system  is the energy of the Dirac
vacuum
$E=-\sum_{p_+}c|p_+|  -\sum_{p_-}c|p_-|$. This energy is
divergent and requires the cut-off, which is provided by the Fermi-momentum
playing the role of the cut-off Planck momentum: $p_F\equiv p_{\rm Planck}$.
Note that even with this cut-off the energy obtained within the effective
theory has a wrong sign, compared with correct microscopic result in
Eq.(\ref{FermiMomentum2}).

The difference between the energies obtained in the microscopic
and the effective theory approaches becomes important if the gravity is
involved, since the energy is the source of the gravitational field. What
kind of the vacuum energy is gravitating is the essence of the cosmological
constant problem.

\subsection{Relevant vacuum energy and cosmological constant}

Inspection of those condensed matter systems,  in which
an effective gravity arises as a low energy phenomenon,
suggests the possible answer: the vacuum energy density
responsible for the cosmological constant is $\tilde
\epsilon =\epsilon-\mu n$ \cite{PhysRepRev,VacuumEnergy}.
This follows from the microscopic  physics:
the conservation of the particle number $N$ requires that
the quantum field theoretical description of the
$N$-body system is given by
${\cal H}-\mu {\cal N}$, where ${\cal
H}$ and ${\cal N}$ are Hamiltonian and particle number
operators in the second quantized form. The energy
$\tilde\epsilon$ does not depend on the choice of the zero
energy level: the shift
$\Delta$ of the zero energy level for one particle leads to
the shift of the chemical potential
$\mu \rightarrow \mu +\Delta$ and of the total energy
$E \rightarrow E +N\Delta$, while   $\tilde E=E-\mu N$
remains invariant.  In terms of $\tilde\epsilon$ the
equation of state of the quantum vacuum  is always
\begin{equation}
P=-\tilde \epsilon~.
\label{EquationState2}
\end{equation}
Though this is obtained using the microscopic
theory ($\tilde \epsilon$ is not determined within the
effective theory), the result does not depend on details of
the quantum liquid: it follows from the thermodynamic
identity in Eq.(\ref{ThermodIdentity}).

The Eq.(\ref{EquationState2}) is the same as the equation of
state of the vacuum in quantum field theory, which follows
from the Einstein cosmological term. Thus $\tilde \epsilon$
serves as the cosmological constant in the effective
gravitational theory. For our vacuum represented by the
Fermi gas, this cosmological constant is big, being
determined by the ``Planck'' energy scale, $\tilde
\epsilon\sim -cp_{\rm Planck}^2$. The minus sign is
in agreement with the negative energy of the Dirac vacuum
in effective theory, and, according to
Eq.(\ref{EquationState2}), this corresponds to the positive
vacuum pressure: Fermi gas (and the Dirac vacuum too) can
can be in equilibrium only in the presence of positive
external pressure
$P$.

There are, however, quantum liquids which can exist
without an external pressure. Liquid $^3$He and liquid
$^4$He at $T=0$ are examples. In both of these liquids
there is some analog of gravity which arises in the low
energy corner. Let us consider the ground state of such
quantum liquid, if there is no contact with the
environment. In a complete equilibrium the pressure $P$ in
the liquid must be zero, since there is no external forces
acting on the liquid. Then from the
Eq.(\ref{EquationState2})  one automatically obtains that
for such equilibrium vacuum at $T=0$ the cosmological
constant in the effective gravity is identically zero,
$\tilde\epsilon\equiv 0$, without any fine tuning. This
means that according to the quantum liquid analogy the
stationary equilibrium vacuum is not gravitating (see more
details in
\cite{VacuumEnergy}).

\section{Mesoscopic Casimir force.}

\subsection{Leakage of vacuum through the wall.}

Now let us discuss the Casimir effect -- the change of the vacuum pressure
caused by the finite size effects in the vacuum.   We must
take into account the discreteness of the spectrum  of
bare particles or quasiparticles (depending on which theory
we use, microscopic or effective) in the slab. Let us start
with the microscopic description in terms of bare
particles (atoms). We can use two different boundary
conditions for particles,  which give two kinds of discrete
spectrum
\begin{eqnarray}
~\omega_k=k{\hbar c\pi\over a}~.
\label{LinearSpectrum1}\\
~\omega_k=\left(k+{1\over 2}\right){\hbar c\pi\over a}~.
\label{LinearSpectrum2}
\end{eqnarray}
Eq.(\ref{LinearSpectrum1})  corresponds to the ``classical
spinless'' fermions with Dirichlet boundary conditions. The
Eq.(\ref{LinearSpectrum2}) is for the 1+1 Dirac fermions
with no particle current through the wall; this case with
the generalization to the
$d+1$ fermions has been discussed in \cite{Paola}.

The vacuum is again represented by the ground state of the
collection of the
$N$ noninteracting particles. We know the structure of the ``vacuum''
completely and thus the ``vacuum energy'' in the slab is well defined: it
is the energy of  $N$ fermions in 1D box of size $a$
\begin{eqnarray}
E(N,a)=\sum_{k=1}^N\omega_k={\hbar c\pi\over 2a}N(N+1)
~~,~{\rm for}~~\omega_k=k{\hbar c\pi\over a}~,
\label{TotalEnergy1}\\
E(N,a)=\sum_{k=0}^{N-1}\omega_k={\hbar c\pi\over 2a}N^2
~~,~~{\rm for}~~\omega_k=\left(k+{1\over 2}\right){\hbar c\pi\over a}~.
\label{TotalEnergy21}
\end{eqnarray}

To calculate the Casimir force acting on the wall, we must introduce the
vacuum on both sides of the wall. Thus let us consider three walls: at
$z=0$, $z=a_1<a$ and $z=a$. Then we have two slabs with
sizes
$a_1$ and $a_2=a-a_1$, and we can find the force acting on the
wall between the two slabs, i.e. at $z=a_1$. We assume
the same Neumann  boundary conditions at all the walls. But
we must allow the particles to transfer between the slabs,
otherwise the main force acting on the wall between the
slabs will be due to the different bulk pressure in the two
slabs.  This can be done due to, say,
a very small holes (tunnel junctions) in the wall, which do
not violate the boundary conditions and do not disturb
the particle energy levels, but still allow the
particle exchange between the two vacua.

This situation can be compared with the traditional Casimir
effect. The force between the conducting plates arises
because the electromagnetic fluctuations of the vacuum in
the slab are modified due to boundary conditions imposed on
electric and magnetic fields. In reality these boundary
conditions are applicable only in the low-frequency limit,
while the wall is transparent for the high-frequency
electromagnetic modes, as well as for the other degrees of
freedom of real vacuum (fermionic and bosonic), that can
easily penetrate through the conducting wall. In the
traditional approach it is assumed that these degrees of
freedom, which produce the divergent terms in the vacuum
energy, must be cancelled by the proper regularization
scheme.  That is why, though the dispersion of dielectic
permeability does weaken the real Casimir force,
nevertheless in the limit of large distances, $a_1\gg
c/\omega_0$, where
$\omega_0$ is the characteristic frequency at which the
dispersion becomes important, the Casimir force does not
depend on how easily the high-energy vacuum leaks through
the conducting wall.

We consider here just the opposite limit, when (almost) all
the bare particles are totally reflected. This corresponds
to the case when the penetration of the high-energy modes of
the vacuum through the conducting wall is highly
suppressed, and thus one must certainly have the
traditional Casimir force.  Nevertheless, we shall show
that due to the mesoscopic finite-$N$ effects the
contribution of the diverging terms to the Casimir effect
becomes dominating. They produce highly oscillating vacuum
pressure in condensed matter. The amplitude of the
mesoscopic fluctuations of the vacuum pressure in this
limit exceeds by factor $p_{\rm Planck}a/\hbar$  the value
of the conventional Casimir pressure. For their description
the continuous effective low-energy theories are not
applicable.

\subsection{Mesoscopic Casimir force in 1d Fermi gas}

The total vacuum energy in two slabs is
\begin{eqnarray}
E(N,a_1,a_2)={\hbar c\pi\over 2}\left({N_1(N_1+1) \over a_1}+{N_2(N_2+1)
\over a_2}\right)~,~\omega_k=k{\hbar c\pi\over a}
\label{TotalEnergyTwoBoxes1}\\
E(N,a_1,a_2)={\hbar c\pi\over 2}\left({N_1^2 \over a_1}+{N_2^2
\over a_2}\right)~,~~\omega_k=\left(k+{1\over 2}\right){\hbar c\pi\over a}~.
\label{TotalEnergyTwoBoxes2}
\end{eqnarray}
\begin{equation}
N_1+N_2=N~,~a_1+a_2=a
\label{ParticleConservation}
\end{equation}
Since particles can transfer between the slabs, the global
vacuum state in this geometry is obtained by minimization
over the discrete particle number
$N_1$ at fixed total number $N$ of particles in the vacuum.
If the mesoscopic
$1/N$ corrections are ignored, one obtains
$N_1\approx(a_1/a)N$ and $N_2\approx (a_2/a)N$, and the force acting on the
wall between the two vacua is zero.

However, $N_1$ and $N_2$ are integer valued, and this leads
to mesoscopic fluctuations of the Casimir force. Within
a certain range of parameter $a_1$ there is a global
minimum characterized by integers ($N_1$,
$N_2$). In the neighboring intervals
of parameters $a_1$, one has either  ($N_1+1$,
$N_2-1$) or ($N_1-1$,
$N_2+1$).  The force
acting on the wall in the state ($N_1$,
$N_2$) is
obtained by variation of $E(N_1,N_2,a_1,a-a_1)$ over
$a_1$ at fixed $N_1$ and $N_2$:
\begin{equation}
F(N_1,N_2,a_1,a_2)=-{dE(N_1,N_2,a_1,a_2)\over
da_1}+{dE(N_1,N_2,a_1,a_2)\over da_2}~.
\label{TotalForce}
\end{equation}
When $a_1$ increases then at some critical
value of $a_1$, where
$E(N_1,N_2,a_1,a_2)=E(N_1+1,N_2-1,a_1,a_2)$, one particle
must cross the wall from the right to the left. At this
critical value the force acting on the wall changes
abruptly (we do not discuss here an interesting physics
arising just at the critical values of
$a_1$, where the degeneracy occurs between the states
($N_1,N_2$)  and  ($N_1+ 1,N_2- 1$); at these positions of
the wall (or membrane) the particle numbers
$N_1$ and
$N_2$ are undetermined and are actually fractional due to
the quantum tunneling between the slabs
\cite{Andreev}). Using for example the spectrum in
Eq.(\ref{TotalEnergyTwoBoxes2}) one obtains for the jump of
the Casimir force:
\begin{equation}
F(N_1\pm 1,N_2\mp 1)-F(N_1,N_2) =\hbar c\pi\left({\pm 2N_1 +1\over
2a_1^2}+{\pm 2N_2-1 \over 2a_2^2}\right)\approx \pm {\hbar c\pi N\over
a_1a_2} ~.
\label{ChaoticForceChange}
\end{equation}
The same result for the amplitude of
the mesoscopic fluctuations is obtained if one uses the spectrum in
Eq.(\ref{TotalEnergyTwoBoxes1}).

In the limit
$a_1\ll a$ the amplitude of the mesoscopic Casimir force
\begin{equation}
|\Delta F_{\rm meso}|= {\hbar c\pi n\over a_1}={\hbar
c\pi n^2\over N_1}\equiv { cp_{Planck}\over a_1}~.
\label{AmplitudeForceChange}
\end{equation}
  It is  by factor $1/N_1=
(\pi\hbar/a_1p_{F})^3\equiv  (\pi\hbar/a_1p_{Planck})^3$
smaller than the vacuum energy density in
Eq.(\ref{FermiMomentum2}). On the other hand it is by the
factor
$p_Fa_1\equiv p_{Planck}a_1$ larger than the
traditional Casimir pressure, which in one-dimensional case
is $P_C\sim  \hbar c/ a_1^2$. The divergent term which
linearly depends on the Planck momentum cutoff
$p_{Planck}$ as in Eq.(\ref{AmplitudeForceChange}) has been
revealed in many different calculations (see e.g.
\cite{CasimirForSphere}),  and attempts have made to invent
the regularization scheme which would cancel the divergent
contribution.

\subsection{Mesoscopic Casimir forces in a general condensed
matter system.}

The equation (\ref{AmplitudeForceChange}) for the
amplitude of the mesoscopic fluctuations of the vacuum
pressure can be immediately  generalized for the
$d$-dimensional space.
The mesoscopic random pressure comes from the discrete
nature of the underlying quantum lquid, which represents
the quantum vacuum. When the volume $V_1$ of the vessel
changes continuously, the equilibium number $N_1$ of
particles changes in step-wise manner. This results in
abrupt changes of pressure at some critical values of the
volume:
\begin{equation}
P_{\rm meso}\sim P(N_1\pm 1)-P(N_1)=\pm {dP\over dN_1}=\pm {mc_s^2\over
V_1}\equiv \pm {cp_{Planck}\over V_1}~,
\label{ChaoticForceChangeGeneral}
\end{equation}
where again $c_s$ is the speed of sound, which plays the
role of the speed of light;  $m$ is mass of atom of the
underlying liquid which plays the role of the Planck
mass.  The mesoscopic pressure is
determined by microscopic ``transPlanckian'' physics, and
thus such microscopic quantity as the mass of the atom, the
``Planck mass'', enters this force.

For the pair correlated systems, such as Fermi
superfluids with finite gap in the energy spectrum, the
amplitude must be twice larger. This is because the jumps
in pressure occurs when two particles (the Cooper pair)
tunnel through the junction,
$\Delta N=\pm 2$. The
transition with $\Delta N=\pm 1$ requires breaking of the
Cooper pair and costs energy equal to the gap.

For the spherical shell of radius $a$
immersed in the quantum liquid the mesoscopic
pressure is
\begin{equation}
P_{\rm meso}\sim  \pm {3mc_s^2\over
4\pi a^3}\equiv
\pm {3p_{\rm Planck}c\over  4\pi  a^3}~,
\label{ChaoticForceChangeGeneralSpherical}
\end{equation}

\section{Discussion.}

Let us compare the mesoscopic vacuum pressure in
Eq.(\ref{ChaoticForceChangeGeneralSpherical}) arising
due to the finite-$N$ effects, with the traditional Casimir
pressure obtained within the effective theories for the same
spherical shell geometry. In case of the original Casimir
effect the effective theory is quantum electrodynamics. In
superfluid
$^4$He this is the low-frequency quantum hydrodynamics,
which is equivalent to the relativistic scalar field
theory.  In other superfluids, in addition to phonons the
other low-energy modes are possible, the massless fermions
and the sound-like collective modes such as spin waves. The
modes with linear (``relativistic'') spectrum in quantum
liquids play the role of the relativistic massless scalar
fields and chiral fermions. The boundary conditions for
the scalar fields are typicallythe Neumann  boundary
conditions, corresponding to the (almost) vanishing mass or
spin current through the wall (let us recall that
there must be some leakage through the
shell to provide the equal bulk pressure on both sides
of the shell).

If we believe in the
traditional  regularization schemes which cancel out the
ultraviolet divergence, then from the  effective scalar
field theory one must obtain the Casimir pressure
$P_C=-dE_C/dV=K
\hbar c_s/8\pi a^4$, where
$K=-0.4439$ for the Neumann boundary conditions;
$K=0.005639$ for the Dirichlet boundary conditions
\cite{CasimirForSphere}. The traditional Casimir pressure is
completely determined by the effective low-energy theory,
it does not depend on the microscopic structure of the
liquid: only the ``speed of light'' $c_s$ enters this
force.  The same pressure will be obtained in case of the
pair-correlated fermionic superflids, if the fermionic
quasiparticles are gapped and their contribution to the
Casimir pressure is exponentially small compared to the
contribution of the massless bosonic collective modes.

However, at least in our case, the result obtained within
the effective theory is not correct: the real Casimir
pressure in  Eq.(\ref{ChaoticForceChangeGeneralSpherical})
is produced by the finite-$N$ mesoscopic effect. It
 essentially depends on the Planck
cut-off parameter, i.e. it cannot be determined by the
effective theory; it is much bigger, by the factor
$p_{\rm Planck}a/\hbar$, than the traditional
Casimir pressure; and it is highly oscillating. The
regularization of these oscillations by, say, averaging
over many measurements; by noise; or due to quantum or
thermal fluctuations of the shell; etc., depend on the
concrete physical conditions of the experiment.

This shows that in some cases the
Casimir vacuum pressure is not within the responsibility
of the effective theory, and the microscopic
(transPlanckian) physics must be evoked. If two systems
have the same low-energy behavior and are described
by the same effective theory, this does not
mean that they necessarily experience the same Casimir
effect. The result depends on many factors, such as the
discrete nature of the quantum vacuum, the ability of
the vacuum to penetrate through the boundaries, dispersion
relation at high frequency, etc. It is not excluded
that even the traditional Casimir effect which comes from
the vacuum fluctuations of the electromagnetic field is
renormalized by the high-energy degrees of freedom

Of course, the extreme limit of almost impenetrable wall,
which we considered, is not applicable to the original
(electromagnetic)  Casimir effect, where the
overwhelming part of the fermionic and bosonic vacua
easily penetrates the conducting walls, and where the
mesoscopic fluctuations must be small. But are they
negligibly small? In any case our example  shows that the
cut-off problem is not the mathematical, but the physical
one, and the physics dictates the proper regularization
scheme or the proper choice of the cut-off parameters.

The dependence of the magnitude of a low-energy effects on
the physics beyond the effective theory was discussed also
in connection with the Chern-Simons term in
relativistic quantum field theories with violated Lorentz
and CPT symmetries
\cite{Jackiw,LorentzCPTviolation}. Quantum liquids provide
an example of the finite high-energy system where the
``transPlanckian'' microscopic physics determines the
coefficient in front of the Chern-Simons term
\cite{PhysRepRev,Volovik1999}, which remains ambiguous
within the effective theory.

I thank A.Yu. Kamenshchik for fruitful discussion.  This
work  was supported in part by the Russian Foundation for
Fundamental Research and by European Science Foundation.

\end{document}